\documentclass[preprint,showpacs,preprintnumbers,amsmath,amssymb]{revtex4}
\usepackage{graphicx}
\begin{document}
\title{Correlation effects and spin dependent transport in carbon nanostructures}
\author{S. Lipi\'nski, D. Krychowski}
\affiliation{%
Institute of Molecular Physics, Polish Academy of Sciences\\M. Smoluchowskiego 17,
60-179 Pozna\'n, Poland
}%
\date{\today}
\begin{abstract}
The impact of symmetry breaking perturbations on the spin dependent transport through carbon nanotube quantum dots in the Kondo regime is discussed.
\end{abstract}

\maketitle

\section{Introduction}

According to many predictions on the evolution  of microprocessor technology \cite{Moore},  around two thousand twenty the limit of a  single atom per bit will be reached  and it is  the range where devices will be   governed by purely  quantum mechanical laws. But before this ultimate atomic limit is reached there is earlier the range of ten atoms per bit and already at this scale silicon technology gets in trouble.  There are many reasons for the predicted collapse of silicon technology,   the most severe limitation comes from thermodynamics. A by-product of computing operation is the excess heat, the relative contribution of  dissipation increases with the scale of integration \cite{Kim}.   Therefore without  improvement of  the energy efficiency, such computers will  just melt during processing.

It is believed   that the next revolution in electronics would be based on molecules \cite{Cuevas} and in particular on carbon systems  \cite{Avouris}.   In contrast to silicon-based devices, the molecular can successfully cope with the more and more demanding miniaturization requirement without worsening their multi-functional properties.

Conventional electronics has  ignored the spin of electron. Recently very rapidly is developing  new branch of nanotechnology - spintronics, which exploits  this degree of freedom. This field brings memory and logic functionalities on the same chip.  Spin - the ultimate logic bit is an  attractive degree of freedom, because the energy scale relevant for its typical dynamics ($10 - 100$ meV)  is order of magnitudes smaller than that involved  in manipulating the electron charge in standard transistors ($1$ eV).  This translates in devices exhibiting ultra-low power consumption and high speed \cite{Naber}. Carbon systems are  particularly attractive  for spintronic applications due to the weak spin-orbit and hyperfine interactions \cite{Cottet}. The corresponding spin-diffusion lengths are of order of several hundreds nms.

In order to move towards quantum information technologies, spintronics at the single spin level is required.  When we connect a single nano-object to two magnetic electrodes, the spin dependent transport is predicted to interplay with single electron physics, i.e. with Coulomb blockade in the case of weak coupling \cite{Sapmaz}  or with the Kondo effect in the case of stronger coupling between the nano-object and electrodes \cite{Kouwenhoven}.

We will present here a very brief review of our recent study of exotic Kondo effect in carbon nanotube quantum dot (CNT-QD) and propositions  how this phenomenon can be exploited in spintronics. Detailed reviews can be found in Refs. \cite{Lipinski1, Lipinski2}.

\section{Spin-orbital Kondo effect}

Fig. 1a   presents schematic view of carbon nanotube quantum dot (CNT-QD).  Such a dot is formed when electrons are confined to a small region within a carbon nanotube \cite{Sapmaz}. Experimentally it is accomplished by laying a CNT  on a silicon dioxide surface, sitting on a doped silicon waver. The silicon wafer serves as the gate electrode.
\begin{figure}[ht]
\includegraphics[width=4 cm,bb=0 0 370 410,clip]{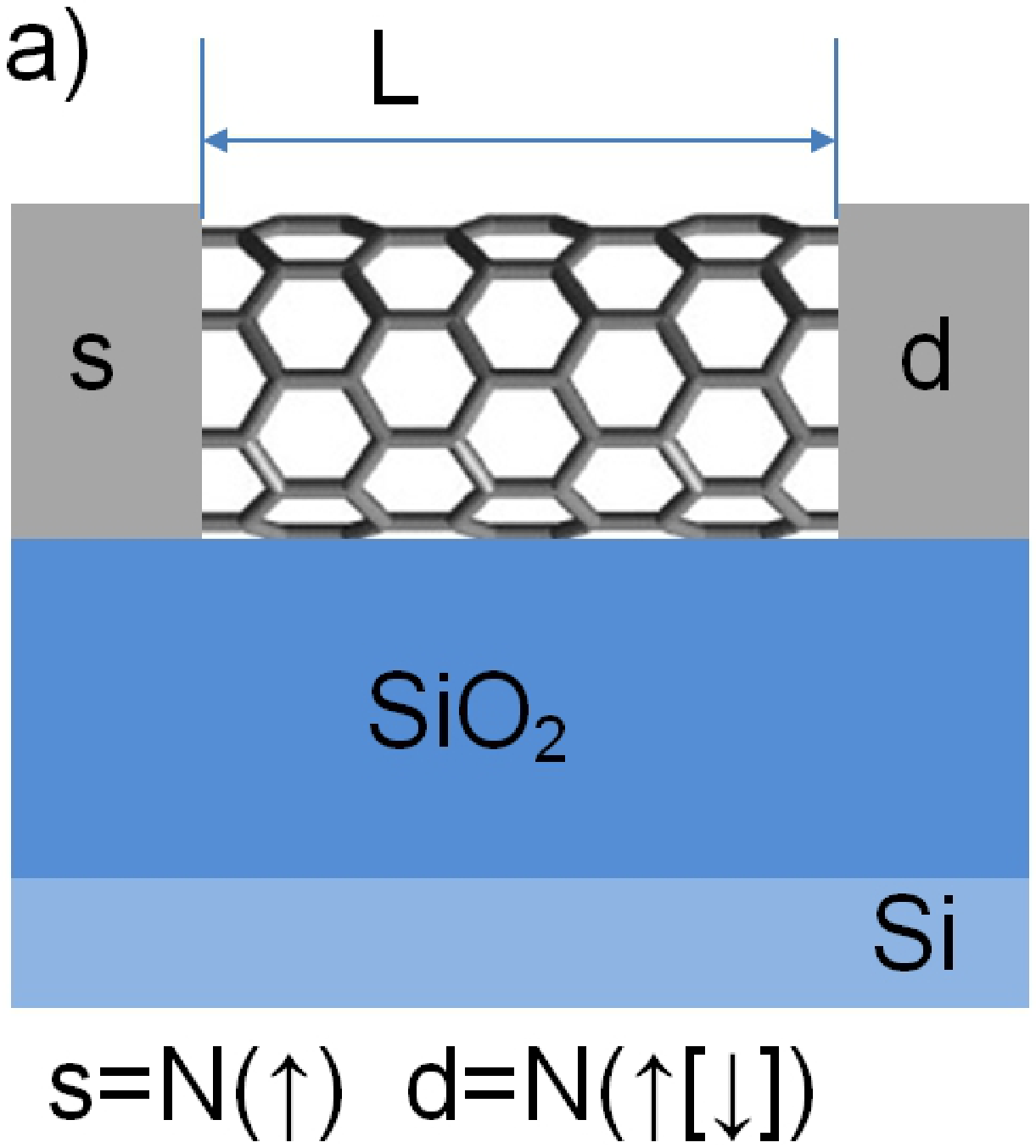}
\includegraphics[width=7 cm,bb=0 0 560 530,clip]{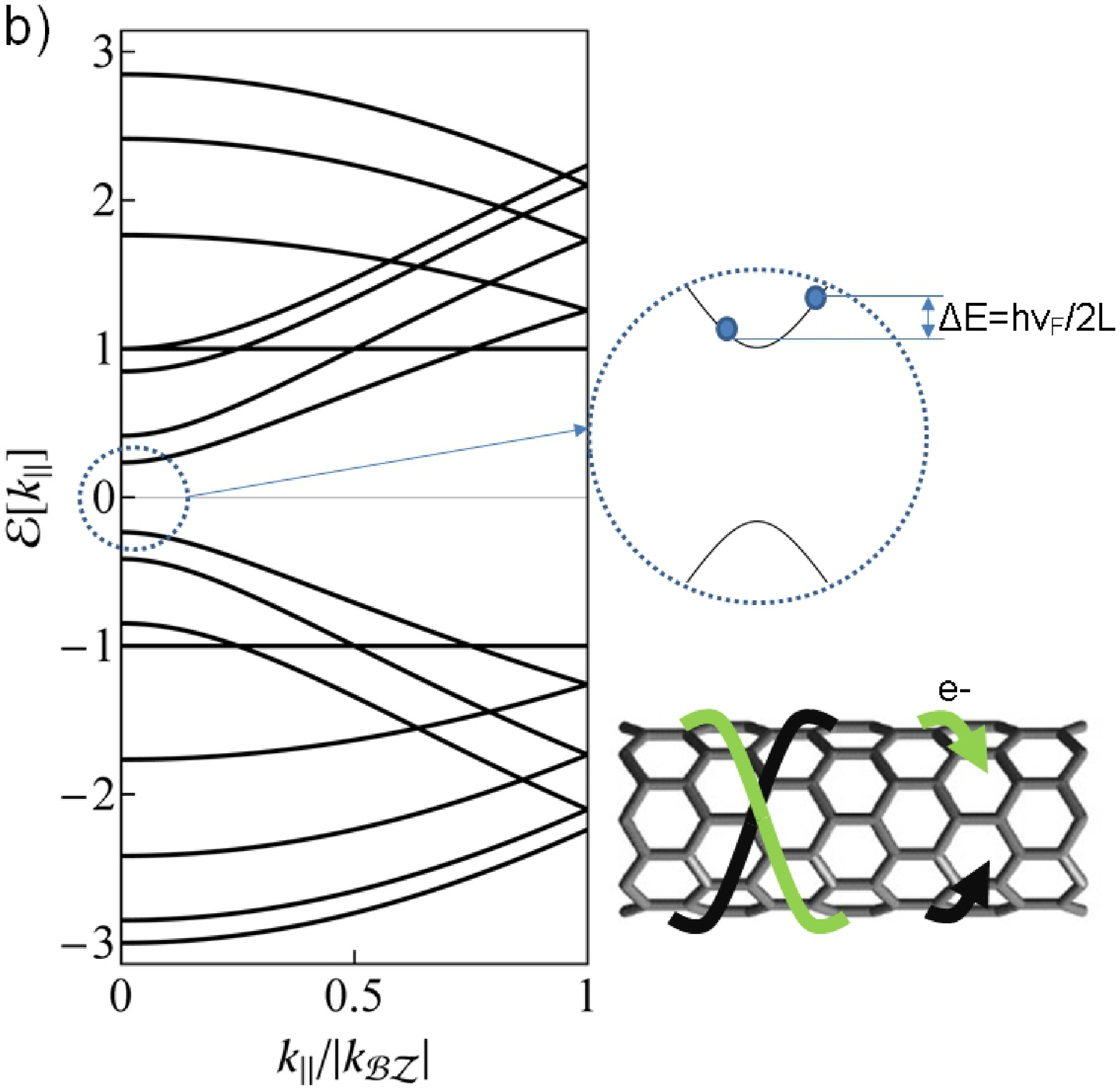}
\caption{a) Side view of CNT-QD with paramagnetic electrodes (N) or with ferromagnetic electrodes (($\uparrow$) or ($\downarrow$) -  spin valve). b) Band structure of semiconducting zigzag CNT C(4,0). Zoom view illustrates quantum confinement quantization of energy levels and  bottom right picture visualizes orbital degeneracy - clockwise and anticlockwise orbital motion.}\label{fig1}
\end{figure}
Metallic leads can then be laid over the nanotube in order to connect the dot up to an electrical circuit. When  the  resistance  of  the  two  tunnel barriers  becomes comparable to or is larger than the quantum resistance (${\cal{R}}_{Q}=h/2e^{2}$), the island becomes strongly separated. A finite length $L$ between the electrodes results in quantized of energy levels  $\Delta E\approx hv_{F}/2L$, where $v_{F}$ denotes Fermi velocity (Fig. 1b).  In an ideal semiconducting nanotube, sets of four electronic states can be grouped together into a shell. All four states are degenerate, with two choices for spin and two for orbital.
\begin{figure}[ht]
\includegraphics[width=8 cm,bb=0 0 660 200,clip]{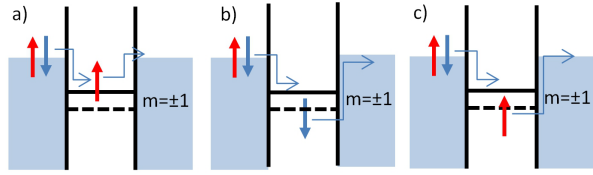}
\caption{Cotunneling processes in SU(4) quantum dot: a) spin-flip fluctuation, b) orbital fluctuation, c) spin-orbital fluctuation.}\label{fig2}
\end{figure}
The orbital degeneracy can be intuitively viewed to originate from two equivalent ways  electrons  can  circle  the  graphene  cylinder,  that  is  clockwise  and  anticlockwise (Fig. 1b).  Kondo effect is observed at low temperatures  in the intermediate coupling range i.e.  when tunnel induced broadening of energy levels $\Gamma$  is smaller than charging energy $\Gamma \leq E_{C}$. In this regime increases the role of higher order tunneling processes. The Kondo spin screening and the associated formation of many-body resonance, which leads  to the enhanced transmission at the Fermi level results from virtual spin flips at the dot caused by tunneling processes off the dot with a given spin followed by tunneling of electron of opposite spin on the dot. Adding  many  spin-flips  processes of  higher  order  coherently,  the  spin-flip  rate  diverges establishing the characteristic temperature - energy scale of quasiparticle resonance.  Kondo temperature sets the  temperature,  respectively  the  voltage  or  magnetic  field  scale  above  which  the  Kondo resonance is suppressed.  The  Kondo  effect  can also occur replacing the spin by orbital \cite{Sasaki} or charge \cite{Wilhelm} degrees of freedom.  The necessary condition for the occurrence of this effect is the same degeneracy of the states in the electrodes and in the QD and conservation of spin or pseudospin in tunneling processes. For the two-fold degenerate states the allowed symmetry operations are rotations in spin space (SU(2)). Spin and orbital degeneracies can also  occur  simultaneously  leading  to  highly  symmetric  Kondo  state  (SU(4)). Most spectacular evidence of this phenomena has been reported by Jarillo-Herrero et al.  for carbon nanotubes \cite{Jarillo}. SU(4)  group characterizes the rotational invariance in spin and orbital space. The simultaneous screening of orbital and spin degrees is caused by tunneling processes causing spin, orbital pseudospin and spin-orbital fluctuations (Fig. 2).
\begin{figure}[ht]
\includegraphics[width=6.8 cm,bb=0 0 660 670,clip]{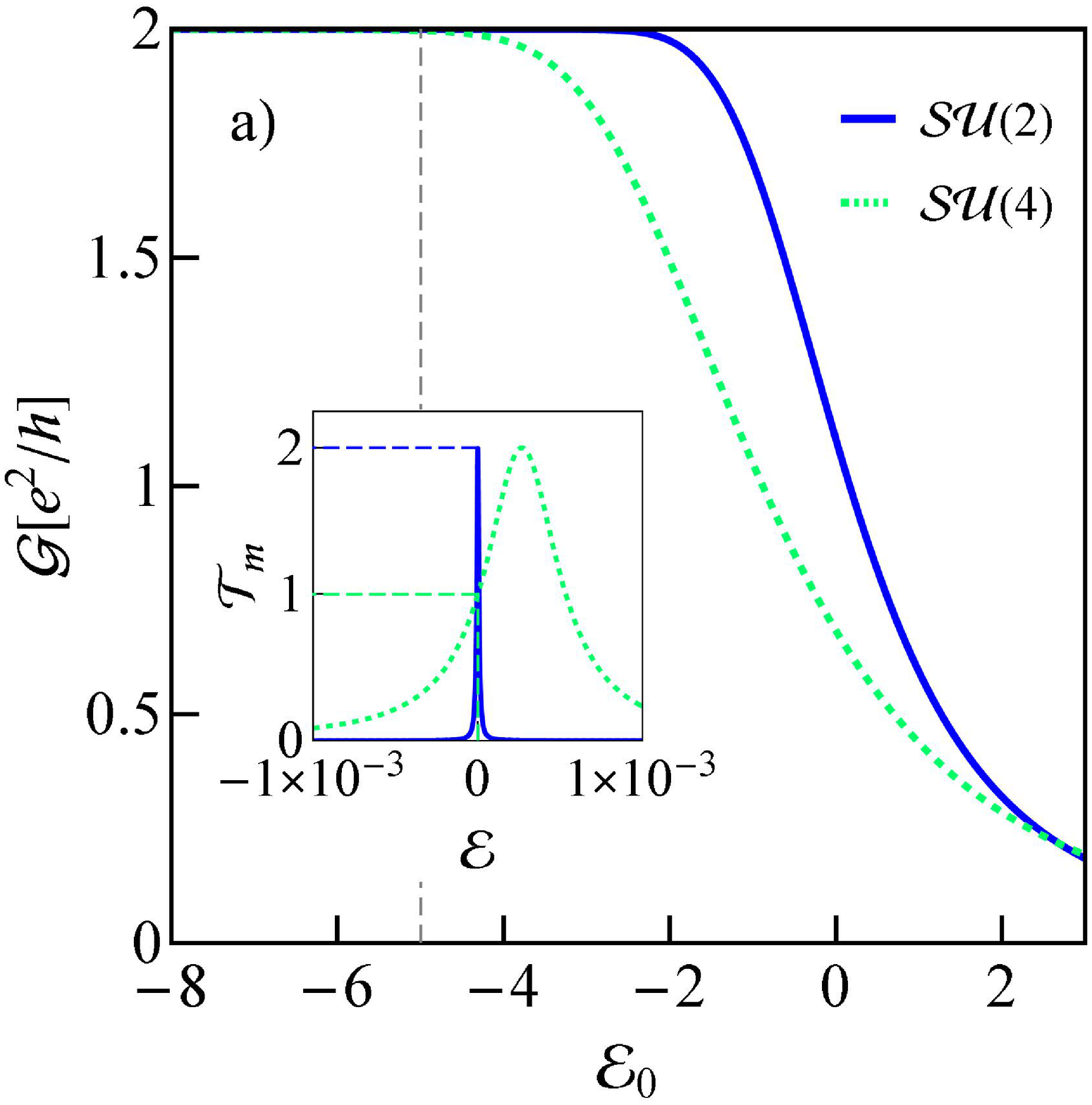}
\includegraphics[width=6.6 cm,bb=0 0 660 704,clip]{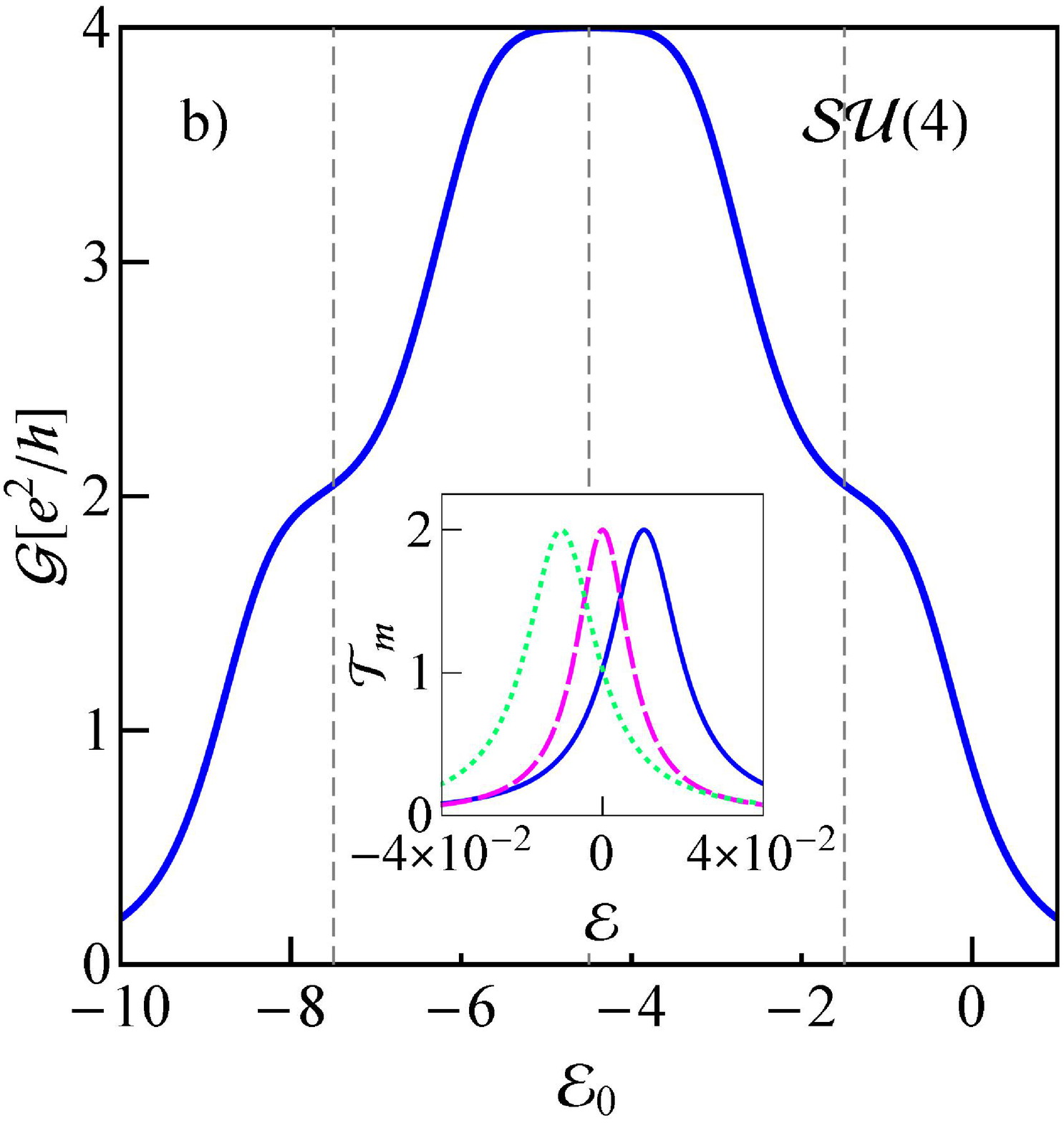}
\caption{a) Linear conductances of  SU(4) and SU(2) Kondo dots vs. dot energy level for $U\rightarrow\infty$. Inset compares corresponding  transmissions for both symmetries ($\varepsilon_{0} = -5$). b) Conductance of SU(4) CNT-QD for $U = 3$. Inset shows the corresponding transmissions from right  to left  for $N = 1,2,3$  valleys (slave boson results - Kotliar-Ruckenstein SBMFA).}\label{fig3}
\end{figure}
In this case orbital psudospins play exactly the same role  as  spins. Fig. 3a compares linear conductances and transmissions for SU(2) and SU(4) symmetries in the infinite  Coulomb interaction limit $U\rightarrow\infty$.  For deep dot level positions the unitary limit of total conductance is reached for both symmetries, and thus from conductance measurements alone one can not distinguish in  the odd electron valleys between these two  symmetries.  The difference can be inferred from measurements in magnetic field or from analysis of shot noise \cite{Lipinski2} or as it is presented in Fig. 3b  by moving to the   two-electron valley range.  The spin-orbital  many body  peak   is  also much broader than spin fluctuation SU(2)  peak,  which means exponential enhancement of Kondo temperature. The use of single wall carbon nanotubes has pushed the Kondo temperatures to the range of several K \cite{Jarillo}, in graphene nanostructures even higher Kondo temperatures are expected \cite{Chen}.  SU(4) transmission  in $N = 1$ valley shows  a  peak  slightly  shifted from  the  Fermi  energy,  it  is  pinned  at  ${\cal{E}}\sim T_{K}$.   For $N = 3$ (single hole) the resonance  is similarly shifted from the Fermi level but  towards negative energy values. We also present in Fig. 3b conductance of CNT-QD in the  Kondo range   for finite U obtained with the use of two-orbital Anderson model. The   pronounced  enhancements of conductance   are observed not only in odd electron valleys, but also in two-electron valley.  In the latter  case all  six degenerate  two electron states participate in the formation of  SU(4) Kondo resonance \cite{Makarovski}. The results are in qualitative agreement with experimental data of Makarovski et al. \cite{Makarovski}, but we have assumed symmetric coupling to the leads, whereas experimental devices were coupled highly asymmetrically.  In the following discussion we will analyze an impact of symmetry breaking perturbations on the spin dependent transport. The numerical results are presented taking  $|e| = g = \mu_{B} = k_{B} = h = 1$, and choosing  $\Gamma$   as the energy unit ($\Gamma = 1$ meV).

\subsection{Kondo spin valve}

The prototype of all spin devices is the spin valve - on which the read heads of computer drives and  magnetic random access memories are based.  It consists of a non-magnetic material sandwiched between two ferromagnetic electrodes. The flow of carriers through a spin valve is determined by the direction of their spin (up or down) relative to the magnetic polarization of the device's electrodes. The relative orientation of polarizations can be tuned by an external magnetic field. The difference in resistance between antiparallel and parallel states is known as its magnetoresistance (MR). In carbon based spintronic device the role of spacer is  played by  graphene, carbon nanotube or fullerene. Fig. 1a can serve as a schematic presentation  of CNT spin valve when electrodes are ferromagnetic.
\begin{figure}[ht]
\includegraphics[width=6.8 cm,bb=0 0 440 450,clip]{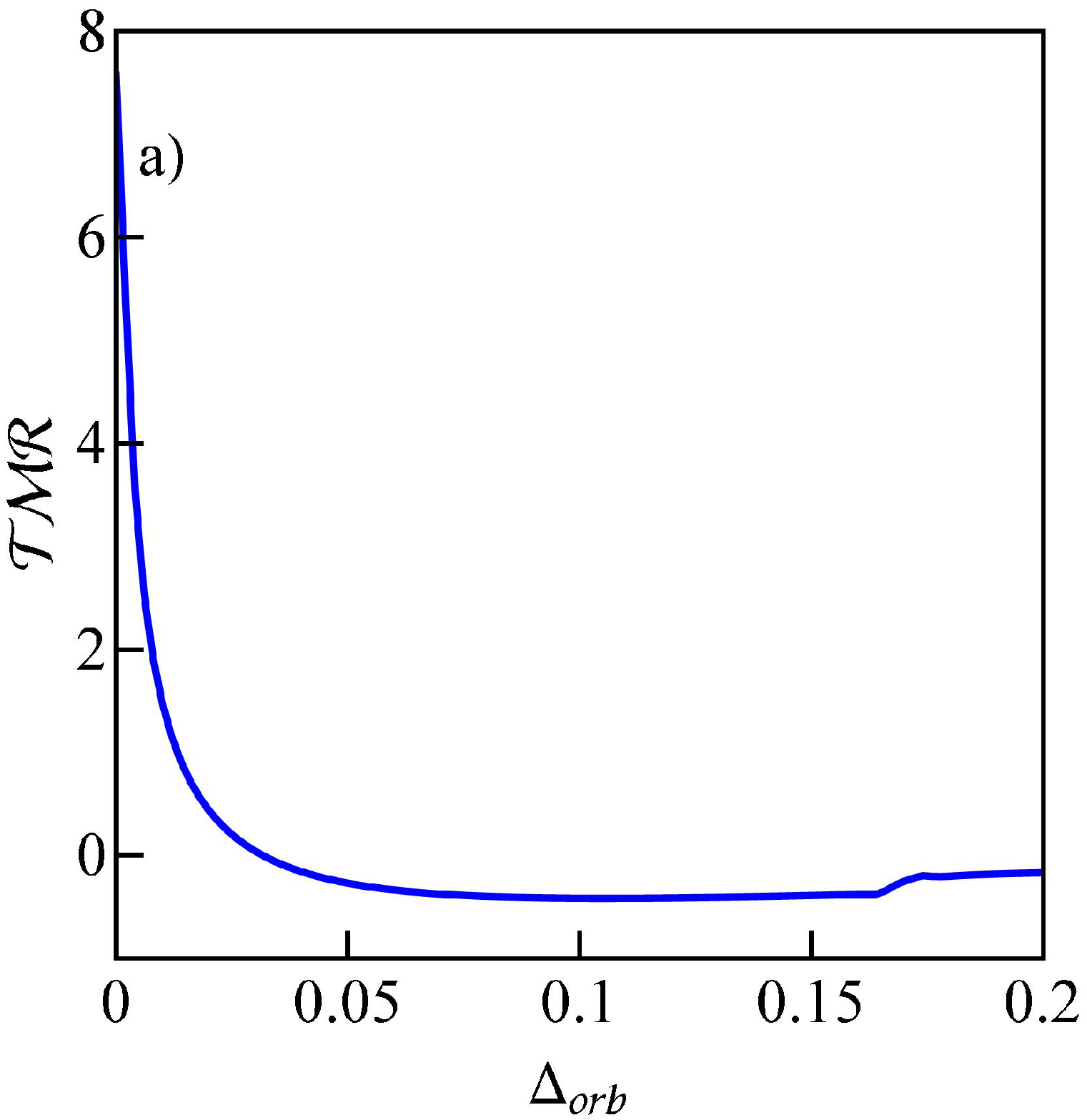}
\includegraphics[width=6.8 cm,bb=0 0 660 680,clip]{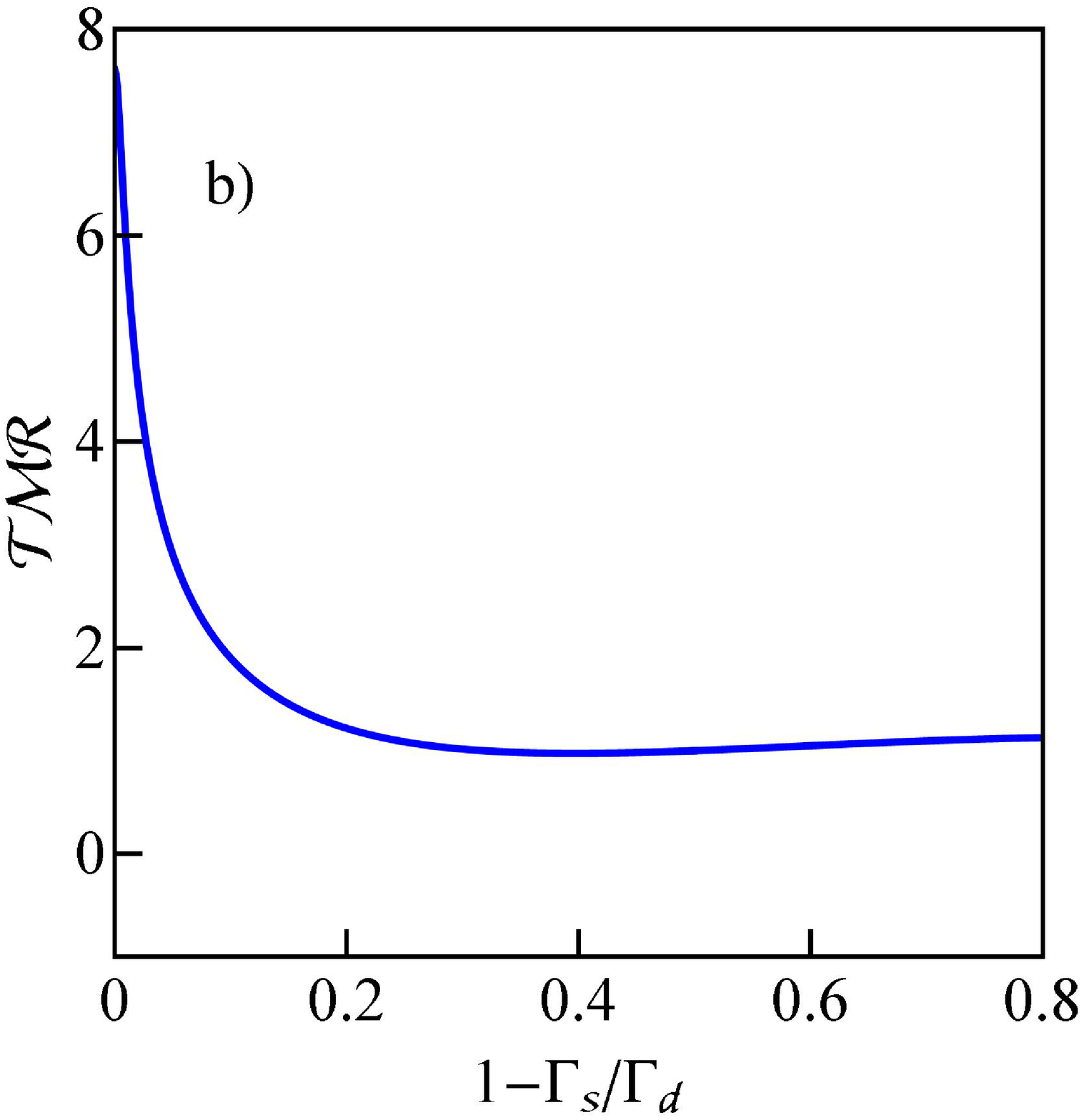}
\caption{Linear TMR as a function of a) orbital level mismatch, b)  asymmetry of couplings ($\Gamma = (\Gamma_{L} + \Gamma_{R})/2$). (Equation of motion (EOM) calculations with  $\varepsilon_{0} = -6$, $U = 15$, polarization of electrodes - relative difference of densities of states for up and down spins at ${\cal{E}}_{F}$, is taken $P = 0.6$).}\label{fig5}
\end{figure}
The first observed  tunneling magnetoresistance  signal (TMR) in molecular system was reported for multiwall carbon nanotube contacted to polycrystalline Co electrodes \cite{Tsukagoshi}.  Recently we have shown, that for  CNT-QDs coupled to ferromagnetic electrodes giant values of TMR are expected in Kondo regime for some ranges of gate voltages \cite{Lipinski2}. The role of polarization of electrodes is twofold: it makes the tunneling processes spin dependent and it introduces an effective exchange field via spin-dependent charge fluctuations \cite{Martinek}. By changing the gate voltage  one can move the system to different ranges of charge fluctuations,  what reflects in the change not only of the value, but also of the sign of effective exchange splitting. This  in turn results in drastic changes of TMR.  Control by electric means  is currently an important challenge for spintronics, because in contrast to an applied magnetic field, electric field  acts rapidly and allows very  localized  addressing. The question arises how robust is TMR against disturbances. Due to disorder or deformations the full spin-orbital degeneracy of  nanotube dots  is rather an  exception than the rule. Two examples of the impact of nonmagnetic perturbations on TMR are shown in Fig. 4, the effect of orbital level mismatch $\Delta_{orb}=\varepsilon_{2} - \varepsilon_{1}$, and influence of asymmetry of coupling of the dot to the electrodes.  It is seen that the record values of magnetoresistance are achieved  for the full symmetric case.

\subsection{Kondo spin filter}

Spin filter  is  a  device  that  filters  electrons  by  their  spin  orientation. In quantum information technology spin filters can be used for initialization and readout of spin quantum bits \cite{Loss}.
\begin{figure}[ht]
\includegraphics[width=6.6 cm,bb=0 0 660 600,clip]{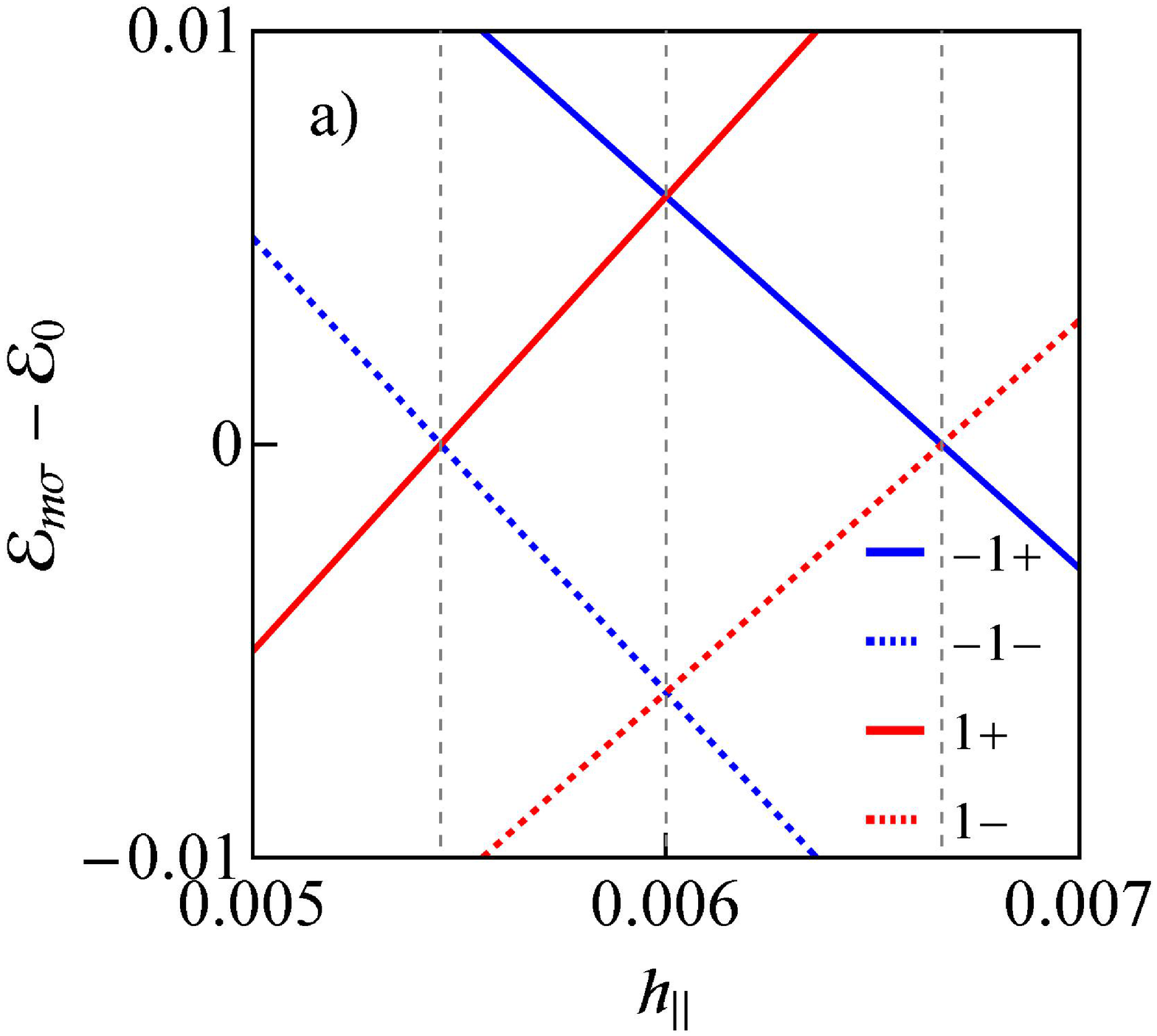}
\includegraphics[width=6.8 cm,bb=0 0 660 600,clip]{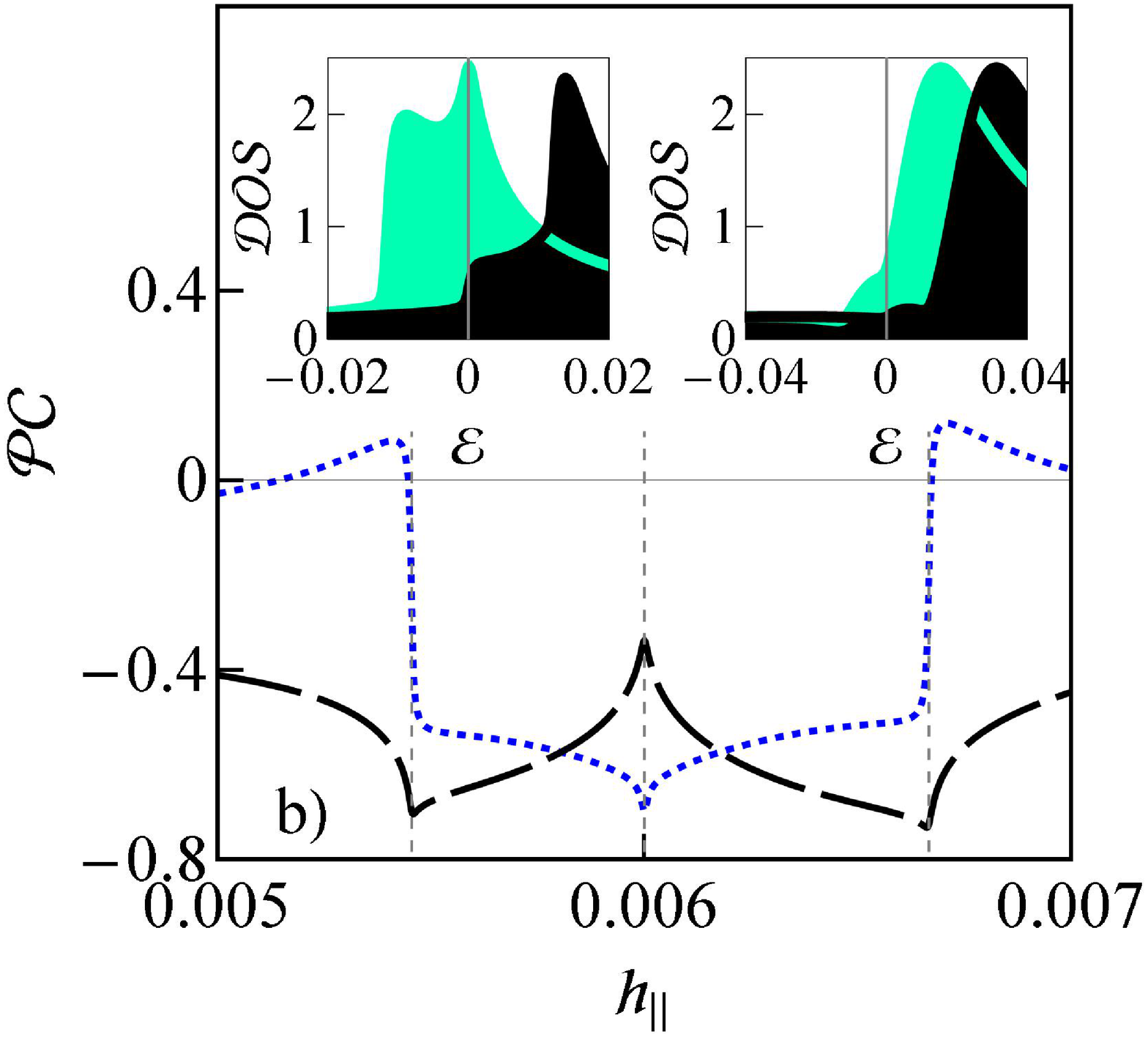}
\caption{a) Illustration of recovery of orbital degeneracy by axial magnetic field ($\Delta_{orb} = 0.012$, $\mu_{orb} = 10 \mu_{B}$).  b) Polarization of conductance vs. axial magnetic field: dotted line is for Kondo regime ($\varepsilon_{0} = -6$) and broken line corresponds to MV range ($\varepsilon_{0} = -4$). Insets show the spin resolved densities of states for $h_{0}$, left for Kondo regime and right for MV case. Dark and grey/green filled curves  correspond to spin up and spin down respectively. (EOM results with $U = 15$).}\label{fig5}
\end{figure}
For systems with broken symmetry  the increase of conductance can be achieved by field induced recovery of degeneracy allowing for  restoration of Kondo effect. Here we discuss tuning of spin polarized orbitally  mismatched states by magnetic field.  If the states are chosen from the same energy shell the filter would  operate in the low  field range, for orbital mismatch  $\Delta_{orb} = 0.01$ meV   the required filtering  fields are of order of $10$ mT.  Orbital level mismatch occurs  in nanotubes with torsional deformation. Axial magnetic field might recover the orbital degeneracy either within the same spin sector ($h_{0} = \frac{\Delta_{orb}}{2\mu_{orb}}$, $\mu_{orb}$ - orbital magnetic moment, typically  $\mu_{orb}\sim10\mu_{B}$)   or with mixing the spin channels ($h_{1,2} = \frac{\Delta_{orb}}{2(\mu_{orb}\mp 1)}$) (see Fig. 5a).  In the former case spin up and spin down  Kondo resonances emerge due to  orbital fluctuations  and in the latter, single Kondo resonance is formed due to spin orbital fluctuations. Apart from the many body resonances resulting from fluctuations between degenerate states  also visible are  satellites reflecting  fluctuations between nondegenerate spin-orbital states (insets of Fig. 5b).  We illustrate the field dependence of polarizations for two representative  gate voltages corresponding to Kondo range of the dot and to the mixed valence range (MV). The  Kondo like  resonances in the Kondo regime are pinned at the Fermi level, whereas in MV state they are shifted above (Fig. 5b). The absolute value of polarization is maximized at $h_{0}$ in the Kondo range when  Kondo peaks for both spin channels cross the Fermi level. In the  MV regime not the maxima but, the tails of the resonances cross ${\cal{E}}_{F}$ in this field. The observed  gate dependence of polarization for the fixed magnetic field opens a path for electric field control of spin polarization.

\subsection{Kondo spin battery}

Recently, there has been an increasing interest in generation of pure spin current without an accompanying charge current \cite{Brataas}. The idea is that when spin-up electrons move to one direction while an equal number of spin-down electrons move to the opposite direction, the net charge current ${\cal{I}}^{{\cal{C}}} = e({\cal{I}}_{+} + {\cal{I}}_{-})$ vanishes and a finite spin current ${\cal{I}}^{{\cal{S}}} = h/2({\cal{I}}_{+} -{\cal{I}}_{-})$ emerges. To be able generate and control spin currents is one of the challenges of spintronics.
\begin{figure}[ht]
\includegraphics[width=6.8 cm,bb=0 0 660 600,clip]{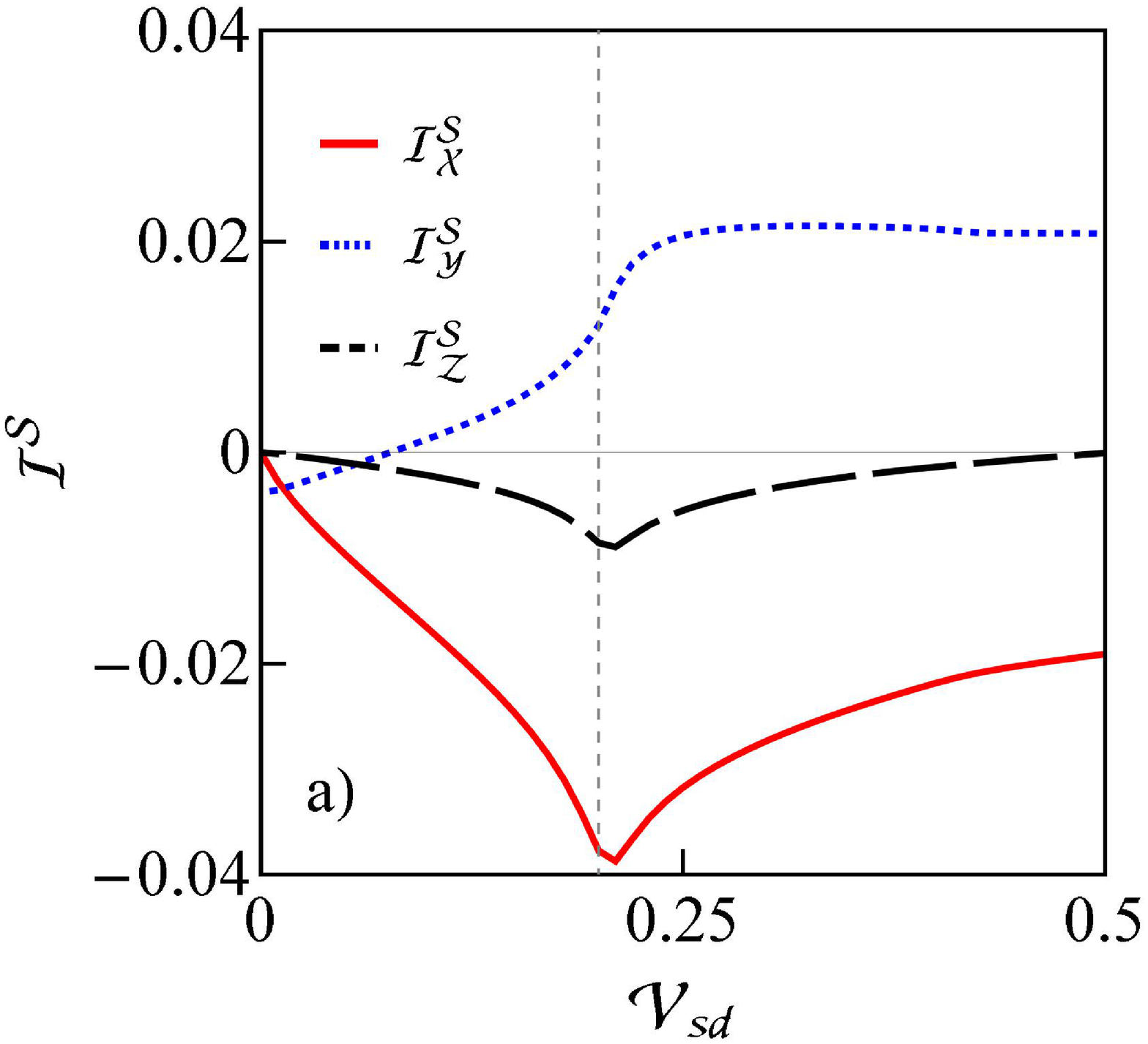}
\includegraphics[width=6.6 cm,bb=0 0 660 610,clip]{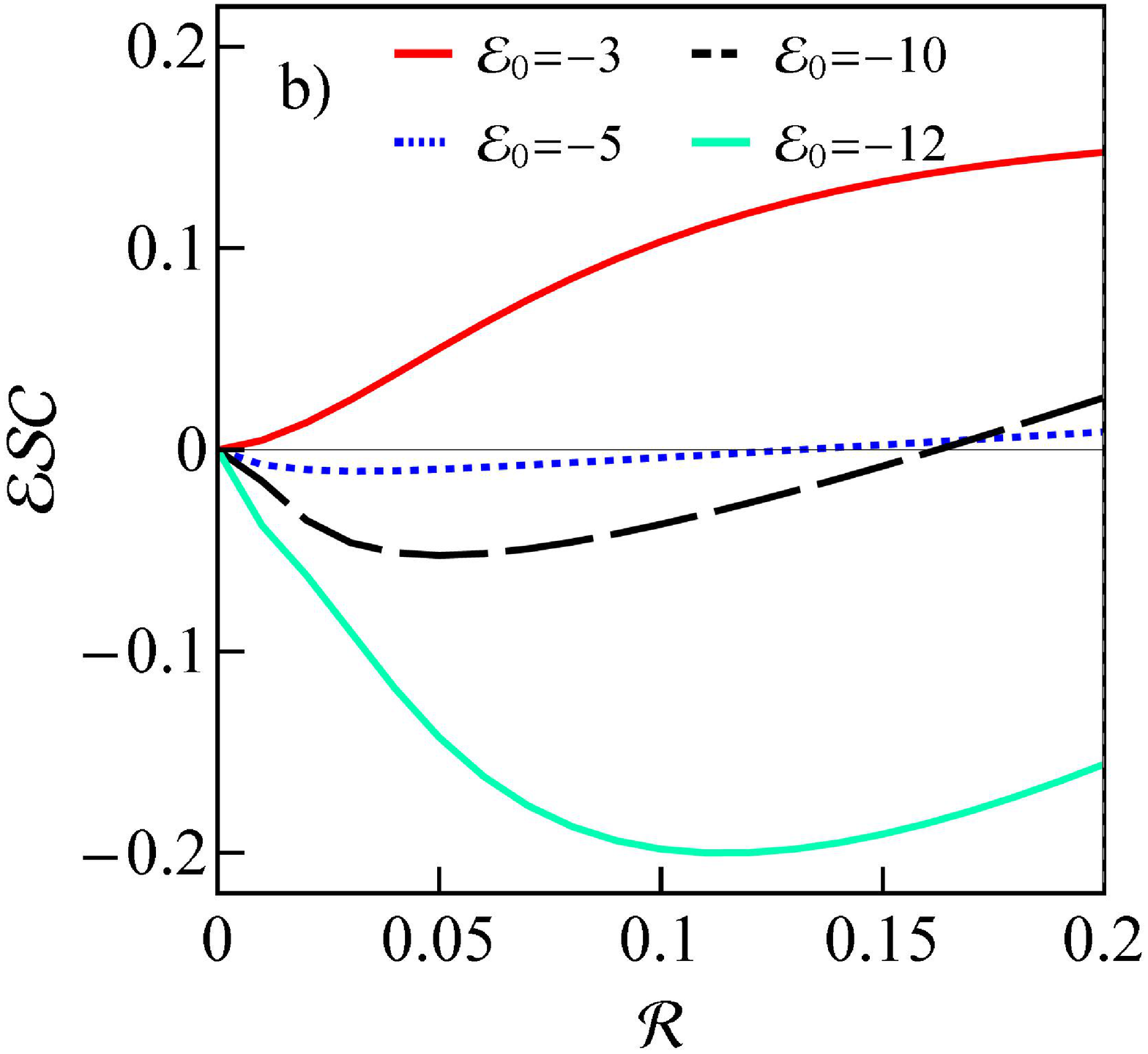}
\caption{a) Spin currents of CNT-QD for spin flip amplitude ${\cal{R}} = 0.1$. b) Equilibrium spin current ${\cal{I}}^{{\cal{S}}}_{y}$ vs.  spin-flip amplitude for several values of dot level (EOM, $U = 15$, $P =0.6$).}\label{fig6}
\end{figure}
The attractive attribute of spin current is that it is associated with a flow of angular momentum, which is a vector quantity. This feature allows information to be sent across nanoscopic structures. As opposed to charge current a spin current is invariant under  time  reversal. This  property  determines  the  low  dissipative  or  even  dissipativeless spin transport  The present proposal presents gate controllable spin-battery based on CNT-QD coupled to ferromagnetic electrodes with different magnetizations. In the following we consider   antiparallely (AP) aligned ferromagnetic leads as illustrated in Fig. 1a.  To allow the device to generate equilibrium spin currents we introduce real spin-flip processes at the dot which mix the spin channels  They are represented by a perturbation ${\cal{H}}'=\sum_{m=1(2)}{\cal{R}}(d_{m+}^{+}d_{m-}+h.c)$.
Spin  flips  may  be  caused  e.g.,  by  transverse  component  of  a  local  magnetic  field. These processes are assumed to be coherent, in the sense that spin-flip strength ${\cal{R}}$ involves reversible transitions. The impact of Kondo effect on the spin currents is of special  interest since it provides also  spin-flip processes (cotunneling mechanism). Real spin flip perturbation  mixes the spin channels and therefore beyond the longitudinal ${\cal{I}}^{{\cal{S}}}_{z}$  spin  current also transverse currents appear for finite bias (Fig. 6a).  But even for vanishing  bias not all components of spin current vanish.   Perturbation ${\cal{H}}'$ is equivalent to operation of magnetic field in $x$ direction and thus the spin torque acting on spin aligned along $z$ direction acts along $y$ direction.  Due to the  opposite spin polarizations of the electrodes  imbalance between the  spin orientations in $y$  direction  for left and right moving carriers is induced. This results in  equilibrium ${\cal{I}}^{{\cal{S}}}_{y}$  component of  spin current ($ESC$).  As it is seen from Fig. 6b  $ESC$ is controllable by gate voltage (different values of $\varepsilon_{0}$)  or spin-flip amplitude (magnetic field), both these factors influence Kondo resonance. For large values of ${\cal{R}}$ the three peak structure of many-body resonances is expected for AP configuration with satellites located roughly at ${\cal{E}}\approx T_{K}\pm2{\cal{R}}$. The satellites  reflect in bias dependence of spin currents as maxima, minima or rapid changes for corresponding voltages (Fig. 6a). The described system would also generate $ESC$ if only one ferromagnetic electrode is attached to the dot with spin flips and such a system can be viewed as spin battery \cite{Brataas}.
\newline
\newline
This work was supported by the Polish Ministry of Science and Higher Education as a research Project No. N N202 199239 for years 2010-2013.

\end{document}